\documentclass[12pt]{spieman}  
\usepackage{amsmath,amsfonts,amssymb}
\usepackage{graphicx}
\usepackage{setspace}
\usepackage{tocloft}
\usepackage{lineno}
\usepackage{algorithm}
\usepackage{algpseudocode}
\usepackage{pgffor}
\usepackage{subcaption}
\title{Constructing efficient channels for ideal observers using the conjugate gradient method}

\author[a,b,*]{Weimin Zhou}
\affil[a]{University of Arizona, Wyant College of Optical Sciences, Tucson, AZ 85721, USA}
\affil[b]{University of Arizona, Department of Radiology \& Imaging Sciences, Tucson, AZ 85721, USA}

\cftpagenumbersoff{figure}
\cftpagenumbersoff{table} 
\begin{document} 
\maketitle

\begin{abstract}

\vspace{2ex}\noindent\textbf{Purpose}: Task-based assessment of image quality (IQ) is critically important for the design and optimization of medical imaging systems. Ideal observers, including the Bayesian Ideal Observer (IO) and the ideal linear observer, i.e., the Hotelling observer (HO), provide objective figures of merit (FOMs) that quantify system performance on signal detection tasks. However, the application of ideal observers to high-dimensional image data is often computationally intractable. Channel mechanisms provide an effective framework for dimensionality reduction that can facilitate the computation of ideal observers. This work presents a conjugate gradient (CG)-based method to construct efficient channels for approximating the IO and HO performance.

\vspace{2ex}\noindent\textbf{Approach}: The channels are constructed using the conjugate gradient (CG) method associated with the optimization of linear discriminant for signal detection tasks. The proposed method is capable of incorporating prior information about the objects being imaged. Numerical studies are conducted for a binary signal detection task involving Type-I lumpy backgrounds. 
The proposed CG-based channels are compared with partial least squares (PLS) channels in terms of their ability to preserve task-relevant information for approximating the IO and HO performance.
The effects of channel number and training dataset size on signal detection performance are systematically investigated.

\vspace{2ex}\noindent\textbf{Results}: For the considered signal-known-exactly (SKE) and background-known-statistically (BKS) binary detection task, the CG-based channels provide efficient low-dimensional representations of images that preserve task-relevant information and enable accurate estimation of the IO and HO performance. Compared with the PLS channels, the proposed method that incorporates prior information about the objects being imaged produces cleaner channels and improved detection performance, particularly under limited training data conditions.

\vspace{2ex}\noindent\textbf{Conclusion}: The proposed CG-based channelization approach provides a principled and efficient way for extracting task-relevant features that enable accurate approximation of the IO and HO performance. By incorporating prior information about the objects being imaged, the proposed method can effectively construct efficient channels using a limited training dataset.

\end{abstract}


{\noindent \footnotesize\textbf{*}Weimin Zhou,  \linkable{weiminzhou@arizona.edu} }

\begin{spacing}{1}   

\section{Introduction}
\label{sect:intro}
Task-based assessment of image quality (IQ) is critically important for the development, evaluation, and optimization of medical imaging systems \cite{barrett2013foundations, barrett1990objective, barrett2015task, reiser2010task, myers2023state}. Such an assessment objectively quantifies the usefulness of the images to an observer for performing specific clinically relevant tasks (e.g., tumor detection). 
Various model observers \cite{barrett1993model, barrett2013foundations ,barrett1998stabilized}, including ideal observers and anthroporphic observers, have been employed and investigated for task-based IQ assessment. The Bayesian Ideal Observer (IO) \cite{barrett1993model,barrett2013foundations} has been widely advocated for task-based assessment of IQ because it provides an upper bound on task performance and quantifies the maximum amount of task-relevant information that can be extracted from image data.
Accordingly,
when the objective of imaging system design is to maximize the amount of task-relevant information, IO performance should be used as a figure-of-merit (FOM).
The IO performance has also been used to benchmark other observers, such as human observers \cite{burgess1981efficiency, myers1985effect, rolland1991ideal}.

When binary signal detection tasks are considered, the IO employs likelihood ratio as its test statistic for decision making \cite{barrett2013foundations, kupinski2003ideal}. Analytic determination of the IO test statistic requires knowledge of the explicit forms of the underlying likelihood functions, which are often intractable in practice. The Hotelling observer (HO) \cite{barrett1991linear, barrett1998stabilized, sanchez2014task} is the ideal linear observer that employs the optimal linear discriminant and is often utilized as a surrogate for the IO for task-based IQ assessment. 
Despite its analytical linear form, direct computation of the HO test statistic can still be challenging for high-dimensional image data because it requires inversion of a large covariance matrix. Moreover, estimating the HO performance from finite data can cause bias \cite{kupinski2007bias}. As such, reducing data dimensionality prior to computing these observers is often desirable.

Recent efforts have explored deep learning–based approaches for estimating the IO performance. In particular, supervised learning methods employing neural networks have been used to learn IO test statistics directly from labeled imaging data \cite{kupinski2002ideal, zhou2019approximating, zhou2020approximating, li2021hybrid}. Additionally, sampling-based approaches using deep generative models have been investigated to approximate the IO test statistic \cite{zhou2023ideal, li2025approximating}. Although these approaches have demonstrated promising performance, they often require large training datasets and substantial computational resources. Consequently, dimensionality reduction remains an important strategy to enable efficient and scalable implementation of these learning-based methods.

Channel mechanisms have been proposed for obtaining low-dimensional representations of images \cite{barrett2013foundations, myers1987addition}. Channel functions can be designed either to represent visual systems for approximating human observers \cite{daugman1988complete, watson1987cortex, yao1992predicting, wilson1979four, abbey2001human, diaz2015derivation} or to efficiently extract task-relevant features for approximating ideal observers \cite{barrett1998stabilized, gallas2003validating, park2007channelized, park2008singular, witten2010partial}. Channels that facilitate the computation of ideal observers are referred to as \textit{efficient channels}. In this work, we focus on the construction of efficient channels.
The IO and HO acting on the channelized data are referred to as channelized IO (CIO) and channelized HO (CHO), respectively. When channels are efficient, the performance of the CIO/CHO approximates the performance of the original IO/HO. 

A variety of channels were explored previously for estimating the IO and HO performance. Barrett \emph{et al.} proposed Laguerre-Gauss (LG) channels that are applicable for signal detection tasks in which the signal to-be-detected is radially symmetric and the correlation in background has no preferred orientation \cite{barrett1998stabilized}.
Park \emph{et al.} proposed singular vector decomposition (SVD) channels that were established based on the known system's point response function \cite{park2008singular}.
However, ideal observers applied with these channels may not represent optimal task performance for more general imaging tasks involving complex object statistics and unknown system response functions.
To address these limitations, data-driven channelization methods have been investigated. For example, 
Witten \emph{et al.} explored the use of partial least squares (PLS) for learning efficient channels directly from image data \cite{witten2010partial}. Granstedt \emph{et al.} recently proposed an autoencoder (AE)-based method for producing efficient channels, in which the AEs were trained to estimate the signal image,  and validated the resulting channels for the HO approximation.  \cite{granstedt2023approximating}. 
More recently, Zhou proposed a method that utilizes gradients of a loss function derived from a Lagrangian associated with the optimal linear discriminant to construct efficient channels for signal detection tasks.
Prior information about objects being imaged can be readily incorporated in this framework, and 
its effectiveness for approximating the HO performance was demonstrated. \cite{zhou2025using}. This approach requires repeatedly inverting the covariance matrices of the channelized data \cite{zhou2025using}.

In this work, we propose an approach to constructing efficient channels based on the conjugate gradient method  \cite{hestenes1952methods, shewchuk1994introduction, greenbaum1997iterative} associated with the optimization of linear discriminant functions for binary  detection tasks.
Specifically, the proposed method constructs orthonormal channels using residuals generated during a CG process that iteratively solves for the Hotelling template.
No matrix inversion is required in this process.
The resulting CG channels are subsequently employed to compute the CHO and CIO, which are systematically validated against the HO performance and the IO performance estimated using a Markov-Chain Monte Carlo (MCMC) method across a range of training conditions.
 A binary detection task was considered in which the signal was modeled as a mixture of Gaussian functions embedded in a lumpy background \cite{rolland1992effect}. The performance of the proposed method was compared with that of the PLS-based channelization method.

The remainder of this article is organized as follows. An overview of model observers and the channel mechanism is provided in Section \ref{sect:bkgd}. The use of the CG method for generating efficient channels is described in Section \ref{sect:method}.  Numerical studies and results are presented in Sections \ref{sect:num} and \ref{sect:result}, respectively.  The paper finally concludes with discussions in Section \ref{sect:conclusion}.

\section{Background}
\label{sect:bkgd}
The data acquisition process of an imaging system is commonly described as:
\begin{equation}
     \mathbf{g} =  \mathcal{H}\mathbf{f}+ \mathbf{n},
\end{equation}
where $\mathbf{g} \in \mathbb{R}^{M}$ denotes the measured image, $\mathcal{H}$ is the imaging operator that maps object $\mathbf{f}$ to image $\mathbf{g}$, and $\mathbf{n} \in \mathbb{R}^{M}$ is the measurement noise.
The object $\mathbf{f}$ should naturally be modeled as a continuous function over space and/or time, and the imaging operator $\mathcal{H}$ should therefore be a continuous-to-discrete (C-D) mapping for a digital imaging system. In applications involving computer simulations, $\mathbf{f}$ is commonly a $N$-dimensional vector that represents a discretized approximation of the object function, and $\mathcal{H}$ is a discrete-to-discrete (D-D) operator. 

\subsection{Binary detection tasks}
Clinical tasks in medical imaging often involve determining whether a diagnostically relevant signal (e.g., tumor) is present or absent in an image.
The imaging processes under the signal-absent ($H_0$) hypothesis and signal-present ($H_1$) hypothesis can be described as:
\begin{equation}\label{eq:cpt2_binary}
\begin{split}
&H_{0}:  \mathbf{g} =  \mathcal{H}\mathbf{f}_b+ \mathbf{n} \equiv \mathbf{b} + \mathbf{n}, \\
&H_{1}:  \mathbf{g} =  \mathcal{H}(\mathbf{f}_b + \mathbf{f}_s)+ \mathbf{n} \equiv \mathbf{b} + \mathbf{s} + \mathbf{n}. 
\end{split}
\end{equation}
Here, $\mathbf{f}_b$ and $\mathbf{f}_s$ denote the functions or vectors that describe the object properties associated with the background and signal, respectively, and $\mathbf{b} \equiv \mathcal{H}\mathbf{f}_b$ and $\mathbf{s} \equiv \mathcal{H}\mathbf{f}_s$ denote the corresponding background image and signal image. Both $\mathbf{f}_b$ and $\mathbf{f}_s$ can be either deterministic or stochastic, depending on the specification of the signal detection task. When the background $\mathbf{f}_b$ is deterministic, the signal detection task is considered background-known-exactly (BKE); while if $\mathbf{f}_b$ is stochastic, the signal detection task is considered background-known-statistically (BKS). Similarly, the signal detection task can be signal-known-exactly (SKE) or signal-known-statistically (SKS). 

To perform a binary detection task, a model observer maps the input image $\mathbf{g}$ to a scalar test statistic $t(\mathbf{g})$ via a discriminant function. The test statistic $t(\mathbf{g})$ is subsequently compared to a predetermined decision threshold $\tau$ to make the decision $D$:
\begin{equation}
D(\mathbf{g}) =
\begin{cases} 
H_1 \text{ is true,} & \text{if } t(\mathbf{g}) \geq \tau, \\
H_0 \text{ is true,} & \text{if } t(\mathbf{g}) < \tau.
\end{cases}
\end{equation}
The false-positive fraction (FPF) and true-positive fraction (TPF) can be computed based on the decision outcomes.  
By varying the decision threshold $\tau$, a receiver operating characteristic (ROC) curve \cite{metz1978basic} that depicts the tradeoff between TPF and FPF can be plotted, and the area under the ROC (AUC) can be evaluated to summarize the observer performance \cite{barrett2013foundations}.

\subsection{Bayesian Ideal Observer}
The Bayesian Ideal Observer (IO) employs the optimal decision-making strategy and establishes an upper bound on detection performance among all observers. The IO test statistic is defined as a likelihood ratio:
\begin{equation}
t_\text{IO}(\mathbf{g}) = \frac{\text{pr}(\mathbf{g}|H_1)}{\text{pr}(\mathbf{g}|H_0)},
\end{equation}
where $\text{pr}(\mathbf{g}|H_1)$ and $\text{pr}(\mathbf{g}|H_0)$ are conditional probability density functions (PDFs) of the image $\mathbf{g}$ under hypotheses $H_1$ and $H_0$, respectively. However, except for a few special cases in which Gaussian or other simple PDFs are involved, analytic determination of the likelihood ratio is typically infeasible. To address this limitation, a Markov-Chain Monte Carlo (MCMC) method was proposed to numerically estimate the IO test statistic by rewriting the likelihood ratio as an integral over the posterior distribution of the background \cite{kupinski2003ideal}:
\begin{equation}
\label{eq:mcmc_IO1}
t_\text{IO}(\mathbf{g}) = \int d\mathbf{b} \Lambda_\text{BKE}(\mathbf{g}|\mathbf{b}) \text{pr}(\mathbf{b}|\mathbf{g}, H_0),
\end{equation}
where $\Lambda_\text{BKE}(\mathbf{g}|\mathbf{b})$ is the BKE likelihood ratio conditioned on an exactly-known background and is determined by noise statistics. When background image can be represented by a stochastic object model parameterized by a vector $\pmb{\theta}$, the IO test statistic can be estimated by Monte Carlo integration:
\begin{equation}
\label{eq:mcmc_IO2}
    \hat{t}_\text{IO}(\mathbf{g}) = \frac{1}{J} \sum_{j=1}^J \Lambda_\text{BKE}(\mathbf{g}|\mathbf{b}(\pmb{\theta}^{(j)})),
\end{equation}
where $\pmb{\theta}^{(j)}$ is sampled from the posterior density $\text{pr}(\pmb{\theta} | \mathbf{g}, H_0)$ and MCMC techniques can be used to achieve this \cite{kupinski2003ideal, abbey2008ideal, zhou2023ideal, li2025approximating}.

\subsection{Hotelling observer}
When implementation of the IO is challenging, the Hotelling observer (HO) is commonly employed as a surrogate.
The HO implements the optimal linear discriminant that maximizes the signal-to-noise ratio of the test statistic, denoted by $\text{SNR}_t$, among all linear observers. The $\text{SNR}_t$ is a commonly used metric for quantifying detection performance and is defined as:
$\text{SNR}_t= \frac{\langle t\rangle_1 - \langle t\rangle_0}{\sqrt{\frac{1}{2}\sigma_0^2+\frac{1}{2}\sigma_1^2}}$,
where $\langle t\rangle_j \equiv \mathbb{E}[t \mid H_j]$ and $\sigma^2_j \equiv \mathrm{Var}[t \mid H_j]$ are the mean and variance of the test statistic $t$ under the hypothesis $H_j$ ($j=0,1$).
 The HO test statistic has an analytic form and is defined as:
\begin{equation}
t_\text{HO}(\mathbf{g}) = \mathbf{w}_\text{HO}^T\mathbf{g},
\end{equation}
where $\mathbf{w}_\text{HO}$ is the Hotelling template and has a closed-form expression:
\begin{equation}
\mathbf{w}_\text{HO} = \left[\frac{1}{2}(\mathbf{K}_0 + \mathbf{K}_1)\right]^{-1}\Delta\bar{\bar{\mathbf{g}}}.
\end{equation}
Here, $\mathbf{K}_j \in \mathbb{R}^{M\times M}$ is the covariance matrix of $\mathbf{g}$ under the hypothesis $H_j$ ($j=0, 1$), and $\Delta\bar{\bar{\mathbf{g}}}$ is the mean-difference vector:
\begin{equation}
    \Delta\bar{\bar{\mathbf{g}}} = \langle \mathbf{g} \rangle_1 - \langle \mathbf{g} \rangle_0 = \langle \bar{\mathbf{g}}(\mathbf{f}) \rangle_{\mathbf{f}|H_1} - \langle \bar{\mathbf{g}}(\mathbf{f}) \rangle_{\mathbf{f}|H_0}, \quad \bar{\mathbf{g}}(\mathbf{f}) \equiv \langle \mathbf{g} \rangle_{\mathbf{g}|\mathbf{f}}.
\end{equation}
Covariance matrix decomposition (CMD) can be useful for estimating and inverting the covariance matrix when computing the Hotelling template \cite{barrett2013foundations}:
\begin{equation}\label{eq:K}
\begin{split}
\mathbf{K}_{j} = &\big\langle  \langle [\mathbf{g} - \bar{\mathbf{g}}(\mathbf{f})] [\mathbf{g} - \bar{\mathbf{g}}(\mathbf{f})]^T   \rangle_{\mathbf{g}|\mathbf{f}}  \big\rangle_{\mathbf{f}|H_j}  + \langle  [\bar{\mathbf{g}}(\mathbf{f}) - \bar{\bar{\mathbf{g}} }_j] [\bar{\mathbf{g}}(\mathbf{f}) - \bar{\bar{\mathbf{g}} }_j]^T  \rangle_{\mathbf{f}|H_j} \\
= &\langle{\mathbf{K}}_{\mathbf{n}|\mathbf{f}} \rangle_{\mathbf{f}|H_j}+ \mathbf{K}_{\bar{\mathbf{g}}(\mathbf{f}) | H_j} \\
\equiv & \overline{\mathbf{K}}_{\mathbf{n} |H_j} + \mathbf{K}_{\overline{\mathbf{g}}|H_j},
\end{split}
\end{equation}
where $\overline{\mathbf{K}}_{\mathbf{n}|H_j}$ is the noise covariance matrix averaged over all objects under the hypothesis $H_j$, and $\mathbf{K}_{\overline{\mathbf{g}}|H_j}$ denotes the covariance matrix corresponding to object variability under the hypothesis $H_j$.
Let $\mathbf{K_b}$ and $\mathbf{K_s}$ denote the covariance matrices associated with the background image and the signal image, respectively, the $\mathbf{K}_{\overline{\mathbf{g}}|H_j}$ can be expressed as:
\begin{equation}
\mathbf{K}_{\overline{\mathbf{g}}|H_0} = \mathbf{K_b}, \quad   \mathbf{K}_{\overline{\mathbf{g}}|H_1} = \mathbf{K_b} + \mathbf{K_s}.
\end{equation}
If the signal is nonrandom, $\mathbf{K}_{\overline{\mathbf{g}}|H_1} = \mathbf{K_b}$. The covariance matrix of the noise can either be known or estimated. An approach to estimating $\mathbf{K_b}$ is to employ stochastic object models, such as lumpy object models \cite{rolland1992effect, bochud1999statistical, kupinski2003experimental} and a binary texture model \cite{abbey2008ideal} or, more recently, those learned using deep generative models, such as AmbientGANs \cite{zhou2019learning, zhou2022learning, xu2025ambient, ozbey2025learning}.

\subsection{Channelized ideal observers and efficient channels}
A practical strategy for mitigating the computational burden associated with computing the ideal observers for high-dimensional images is to perform dimensionality reduction through data channelization. If the channels are efficient, the performance of the ideal observers operating on the lower-dimensional channelized data closely approximates the performance of the ideal observers operating on the original image data.
Given a channel matrix $\mathbf{T}\in \mathbb{R}^{D\times M}$ that represents a collection of $D$ ($D \ll M$) channel vectors $\{\mathbf{t}_{i}\}_{i=0}^{D-1}$:
\begin{equation}
    \mathbf{T}=\begin{bmatrix}
\mathbf{t}_0^T \\
\mathbf{t}_1^T\\
\vdots \\
\mathbf{t}_{D-1}^T
\end{bmatrix}, \quad \mathbf{t}_i \in \mathbb{R}^{M\times 1},
\end{equation}
the channelized data of an image $\mathbf{g}$ is: $\mathbf{v} = \mathbf{T}\mathbf{g}$.
The channelized Ideal Observer (CIO) can be subsequently implemented by employing the test statistic given by the likelihood ratio of the channelized data:
\begin{equation}
t_\text{CIO}(\mathbf{g}) = \frac{\text{pr}\left(\mathbf{v}(\mathbf{g}) | H_1\right)}{\text{pr}\left(\mathbf{v}(\mathbf{g}) | H_0\right)}.
\end{equation}
Similar to the MCMC-based IO computation described in Eqs. (\ref{eq:mcmc_IO1}) and (\ref{eq:mcmc_IO2}), the CIO test statistic can also be estimated using MCMC techniques \cite{park2007channelized}. 

The channelized Hotelling observer (CHO), which is the HO operating on the channelized data $\mathbf{v}$, computes its test statistic as:
\begin{equation}
t_{\mathrm{CHO}}(\mathbf{g})
= t_{\mathrm{HO}}(\mathbf{v})
= \mathbf{w}_v^{T} \mathbf{v},
\end{equation}
where $\mathbf{w}_v=\mathbf{K}_v^{-1} \Delta\bar{\bar{\mathbf{v}}}$ is the Hotelling template for channelized data $\mathbf{v}$.
Here,
\begin{equation}
\begin{split}
&\Delta\bar{\bar{\mathbf{v}}} = \bar{\bar{\mathbf{v}}}_1 - \bar{\bar{\mathbf{v}}}_0 = \langle \bar{\mathbf{v}}(\mathbf{f}) \rangle_{\mathbf{f}|H_1} - \langle \bar{\mathbf{v}}(\mathbf{f}) \rangle_{\mathbf{f}|H_0}, \quad \bar{\mathbf{v}}(\mathbf{f}) \equiv \langle \mathbf{v}(\mathbf{g}) \rangle_{\mathbf{g}|\mathbf{f}}. \\
&\mathbf{K}_v = \frac{1}{2} \left[ \mathbf{K}_{v,0} + \mathbf{K}_{v,1} \right], 
\quad 
\mathbf{K}_{v,j} = \mathbb{E}\left[ (\mathbf{v} - \bar{\bar{\mathbf{v}}}_j)(\mathbf{v} - \bar{\bar{\mathbf{v}}}_j)^T \mid H_j \right].
\end{split}
\end{equation}

 Various strategies have been proposed to construct efficient channels. Early work by Barrett \emph{et al.} introduced Laguerre–Gauss (LG) channels, which are formed by combining Gaussian functions with Laguerre polynomials \cite{barrett1998stabilized}. These channels have proven effective for estimating the HO performance for tasks involving rotationally symmetric signals embedded in backgrounds with a correlation that has no preferred orientation, such as lumpy backgrounds. However, they may be suboptimal for more complex backgrounds and non-symmetric signals \cite{granstedt2023approximating}.
Another approach to designing efficient channels is based on singular value decomposition (SVD), where the channels are defined as the singular vectors of the imaging operator \cite{park2008singular}. However, this approach can be difficult to apply in practice, as it relies on explicit knowledge of the imaging operator and requires computing its SVD, which may be computationally expensive for high-dimensional problems.

Data-driven methods have been investigated for generating efficient channels.
For example,
Witten \emph{et al.} investigated the use of the partial least squares (PLS) method, a supervised learning approach for dimensionality reduction, to construct efficient channels \cite{witten2010partial}. This approach generates channels directly from training images without prior knowledge of the system operator or specific assumptions about object statistics.
However, like other supervised learning methods, the performance of PLS channels may depend on the availability of sufficient training samples. 
Granstedt \emph{et al.} proposed an autoencoder (AE)–based method for generating efficient channels for approximating the HO, in which the AE was trained to estimate the signal image \cite{granstedt2023approximating}. It has been shown that incorporating a cleaner estimate of the signal image in the AE-learned channels can lead to improved signal detection performance \cite{granstedt2023approximating}. 

More recently, Zhou proposed utilizing gradients of a Lagrangian function associated with the optimal linear discriminant to generate efficient channels. This framework provides a data-driven approach that permits incorporation of prior information about objects, and its effectiveness for approximating the HO performance has been demonstrated \cite{zhou2025using}. However, this method requires repeatedly inverting covariance matrices of channelized data. In this work, we propose an alternative approach based on the conjugate gradient method for constructing efficient channels without matrix inversion, and evaluate its effectiveness for approximating both the HO and IO performance. 

\section{Conjugate gradient-based efficient channels}
\label{sect:method}
In this work, we propose a conjugate gradient (CG)-based method for constructing efficient channels for signal detection tasks. By interpreting the Hotelling template as the solution to a linear system, the CG process progressively generates orthogonal channels that form an expanding subspace in which increasingly accurate approximations to the optimal linear discriminant are achieved.

\subsection{Conjugate direction method for estimating the Hotelling template}
The CG-based method for constructing efficient channels is based on a quadratic optimization problem associated with the estimation of the Hotelling template.
It can be shown that the optimal linear discriminant for a binary detection task can be obtained by minimizing the following quadratic function:
\begin{equation}
\label{eq:loss}
l(\mathbf{w}) = \frac{1}{2}\mathbf{w}^T\mathbf{K_g}\mathbf{w} -  \Delta \bar{ \bar{ \mathbf{g} } } ^T \mathbf{w}
\end{equation}
For a positive definite covariance matrix $\mathbf{K_g}$, this loss function is minimized by the Hotelling template $\mathbf{w}_\text{HO} = \mathbf{K_g}^{-1}\Delta \bar{\bar{\mathbf{g}}}$.
The method of conjugate directions can solve this problem by iteratively constructing solutions within an expanding Krylov subspace \cite{shewchuk1994introduction}. Specifically, the conjugate direction method identifies a sequence of search directions $\mathbf{d}_{(0)}, \mathbf{d}_{(1)}, ..., \mathbf{d}_{(M-1)}$ that are mutually conjugate, i.e., $\mathbf{K_g}$-orthogonal: 
\begin{equation}
\mathbf{d}_{(i)}\mathbf{K_g}\mathbf{d}_{(j)} = 0, \quad i\neq j,
\end{equation}
and iteratively updates the solution along these directions:
\vspace{-0.5em}
\begin{equation}
\label{eq:w}
\mathbf{w}_{(i+1)} = \mathbf{w}_{(i)} + \alpha_{(i)} \mathbf{d}_{(i)},
\end{equation}
where $\alpha_{(i)}$ is the step size that minimizes the objective function in Eq. (\ref{eq:loss}) evaluated at $\mathbf{w}_{(i+1)}$ and is given by \cite{shewchuk1994introduction}:
\begin{equation}
\label{eq:cd-alpha}
\alpha_{(i)} = \frac{\mathbf{d}_{(i)}^T\mathbf{r}_{(i)}}{\mathbf{d}_{(i)}^T\mathbf{K_g}\mathbf{d}_{(i)}}.
\end{equation}
Here, $\mathbf{r}_{(i)} = \Delta \bar{ \bar{ \mathbf{g} } } - \mathbf{K_g}\mathbf{w}_{(i)}$ is the residual vector that equals the negative gradient of the loss function in Eq. (\ref{eq:loss}).
It can be shown that
\vspace{-0.5em}
\begin{equation}
\label{eq:ris}
\mathbf{r}_{(i+1)} = \mathbf{r}_{(i)} - \alpha_{(i)}\mathbf{K_g}\mathbf{d}_{(i)},
\end{equation}
and $\mathbf{r}_{(i+1)}$ is orthogonal to the subspace spanned by $\{\mathbf{d}_{(0)}, \mathbf{d}_{(1)},..., \mathbf{d}_{(i)}\}$ \cite{shewchuk1994introduction}.
To form a set of $\mathbf{K_g}$-orthogonal search directions $\{\mathbf{d}_{(i)}\}_{i=0}^{M-1}$, a conjugate Gram-Schmidt process can be employed. Specifically, given a set of $M$ linearly independent vectors $\{\mathbf{u}_{(i)}\}_{i=0}^{M-1}$, and set $\mathbf{d}_0 = \mathbf{u}_0$, the $\mathbf{d}_{(i)}$ ($i>0$) is generated by subtracting from $\mathbf{u}_{(i)}$ any components that are not $\mathbf{K_g}$-orthogonal to the previous direction vectors $\mathbf{d}_{(k)}$ ($k=0,...,i-1$) \cite{shewchuk1994introduction}:
\vspace{-0.5em}
\begin{equation}
\label{eq:cd-d}
\mathbf{d}_{(i)} =\mathbf{u}_{(i)} + \sum_{k=0}^{i-1}\beta_{ik}\mathbf{d}_{(k)},
\end{equation}
where $\beta_{ik}$ is the Gram-Schmidt constant and can be determined by the conjugacy condition:
\begin{equation}
\beta_{ik} = -\frac{\mathbf{u}_{(i)}^T\mathbf{K_g}\mathbf{d}_{(k)}}{\mathbf{d}_{(k)}^T\mathbf{K_g}\mathbf{d}_{(k)}}
\end{equation}
Because $\{\mathbf{d}_{(i)}\}_{i=0}^{j-1}$ is constructed from  $\{\mathbf{u}_{(i)}\}_{i=0}^{j-1}$, they span the same subspace.
Furthermore, since the residual $\mathbf{r}_j$ is orthogonal to the subspace spanned by $\{\mathbf{d}_{(i)}\}_{i=0}^{j-1}$ \cite{shewchuk1994introduction}, it follows that:
\begin{equation}   
\label{eq:ru}
\mathbf{r}_{(j)}^T \mathbf{u}_{(i)} =\mathbf{r}_{(j)}^T \mathbf{d}_{(i)}= 0, \quad j>i.
\end{equation}
This orthogonality provides a principled mechanism for constructing mutually orthogonal channels, as will be discussed in the following subsection.

\subsection{Conjugate gradient method for constructing efficient channels}
The conjugate gradients (CG) method is a conjugate direction method that utilizes the residual $\mathbf{r}_{(i)}$ as a vector $\mathbf{u}_{(i)}$ to form the search direction $\mathbf{d}_{(i)}$. By substituting $\mathbf{u}_{(i)}$ with $\mathbf{r}_{(i)}$ in Eq. (\ref{eq:ru}), it follows that the vector $\mathbf{r}_{(j)}$  is orthogonal to the previous residual vectors:
\begin{equation}
\label{eq:r_orth}
\mathbf{r}_{(j)}^T\mathbf{r}_{(i)}=0, \quad j > i.
\end{equation}
This residual vector, which is the negative gradient of the loss function in Eq. (\ref{eq:loss}), will be used as a channel after normalization.
Because the residual vectors are mutually orthogonal, as described in Eq. (\ref{eq:r_orth}), the resulting normalized channels are orthonormal.

The iterative CG process begins with initialization of $\mathbf{w}_{(0)}$ with the residual $\mathbf{r}_{(0)} = \Delta \bar{ \bar{ \mathbf{g} } } - \mathbf{K_g}\mathbf{w}_{(0)}$ and the search direction $\mathbf{d}_{(0)} = \mathbf{r}_{(0)}$. 
Given the residual $\mathbf{r}_{(i)}$ and the search direction $\mathbf{d}_{(i)}$ at the $i$-th iteration, 
the step size $\alpha_{(i)}$ is computed to update the template $\mathbf{w}_{(i+1)}$ according to Eq. (\ref{eq:w}) and to obtain the residual  $\mathbf{r}_{(i+1)}$ according to Eq. (\ref{eq:ris}).
According to Eqs. (\ref{eq:cd-alpha}) and (\ref{eq:cd-d}), together with the orthogonality relation between $\mathbf{r}_{(i)}$ and $\{\mathbf{d}_{(k)}\}_{k=0}^{i-1}$ given in Eq. (\ref{eq:ru}), the step size $\alpha_{(i)}$ can be expressed as: 
\begin{equation}
    \alpha_{(i)} = \frac{\mathbf{r}_{(i)}^T\mathbf{r}_{(i)}}{\mathbf{d}_{(i)}^T\mathbf{K_g}\mathbf{d}_{(i)}}
\end{equation}
Once $\mathbf{r}_{(i+1)}$ is obtained, the search direction can be updated according to the conjugate Gram-Schmit process:
\begin{equation}
\mathbf{d}_{(i+1)} =\mathbf{r}_{(i+1)} + \sum_{k=0}^{i}\beta_{i+1,k}\mathbf{d}_{(k)},
\end{equation}
where the Gram-Schmidt constant $\beta_{i+1,k} = -\frac{\mathbf{r}_{(i+1)}^T\mathbf{K_g}\mathbf{d}_{(k)}}{\mathbf{d}_{(k)}^T\mathbf{K_g}\mathbf{d}_{(k)}}$ and can be simplified as \cite{sanchez2014task}:
\begin{equation}
\beta_{i+1, k} = \begin{cases}
    \frac{1}{\alpha_{(i)}}\frac{\mathbf{r}_{(i+1)}^T\mathbf{r}_{(i+1)}}{\mathbf{d}_{(i)}^T\mathbf{K_g}\mathbf{d}_{(i)}} = \frac{\mathbf{r}_{(i+1)}^T\mathbf{r}_{(i+1)}}{\mathbf{r}_{(i)}^T\mathbf{r}_{(i)}}& \text{if } k = i, \\
    0 & \text{if } k < i.
\end{cases}
\end{equation}
Let $\beta_{(i+1)}$ denote $\beta_{i+1,i}$, the updated conjugate direction $\mathbf{d}_{(i+1)}$ is:
\begin{equation}
\mathbf{d}_{(i+1)} =\mathbf{r}_{(i+1)} + \beta_{(i+1)}\mathbf{d}_{(i)}.
\end{equation}
The subsequent iteration can then be performed. The proposed procedure for generating the CG-based channels is summarized in Algorithm \ref{alg}:

\begin{algorithm}[H]
\caption{CG process for generating efficient channels}
\small
\begin{algorithmic}[1]
\Require Covariance matrix $\mathbf{K_g}$, mean-difference vector $\Delta \bar{ \bar{ \mathbf{g} } }$, number of channels $D$,  threshold $\tau_{\mathrm{orth}}$
\State Initialize $\mathbf{w}_{(0)} = 0$, $\mathbf{r}_{(0)} = \Delta \bar{ \bar{ \mathbf{g} } } - \mathbf{K_g} \mathbf{w}_{(0)}$, $\mathbf{d}_{(0)} = \mathbf{r}_{(0)}$, $\mathbf{t}_0 = \frac{\mathbf{r}_{(0)}}{\| \mathbf{r}_{(0)}\|}$
\For{$i = 0, \dots, D-1$}
    \State {Update step size} $\alpha_{(i)} = \frac{\mathbf{r}_{(i)}^T \mathbf{r}_{(i)}}{\mathbf{d}_{(i)}^T \mathbf{K}_g \mathbf{d}_{(i)}}$ 
    \State {Update template} $\mathbf{w}_{(i+1)} = \mathbf{w}_{(i)} + \alpha_{(i)} \mathbf{d}_{(i)}$
    \State {Update residual} $\mathbf{r}_{(i+1)} = \mathbf{r}_{(i)} - \alpha_{(i)} \mathbf{K_g} \mathbf{d}_{(i)}$
   \If {$\max\limits_{j=0,...,i}\frac{|\mathbf{r}_{(j)}^T\mathbf{r}_{(i+1)} |}{\|\mathbf{r}_{(j)} \| \|\mathbf{r}_{(i+1)} \|} > \tau_{orth}$}, use the Modified Gram-Schmidt for re-orthogonalization:
        \For{$j = 0,\dots,i$}
            \State $\mathbf{r}_{(i+1)} \leftarrow \mathbf{r}_{(i+1)} - \dfrac{\mathbf{r}_{(j)}^T\mathbf{r}_{(i+1)}}{\| \mathbf{r}_{(j)}\|^2}\mathbf{r}_{(j)}$

        \EndFor
   \EndIf
   \State {Produce normalized channel} $\mathbf{t}_{{i+1}}=\frac{\mathbf{r}_{{(i+1)}}}{\| \mathbf{r}_{(i+1)}\|}$
    \State {Update the Gram-Schmidt constant} $\beta_{(i+1)} = \frac{\mathbf{r}_{(i+1)}^T \mathbf{r}_{(i+1)}}{\mathbf{r}_{(i)}^T \mathbf{r}_{(i)}}$
    \State {Update the search direction} $\mathbf{d}_{(i+1)} = \mathbf{r}_{(i+1)} + \beta_{(i+1)} \mathbf{d}_{(i)}$
\EndFor
\end{algorithmic}
\label{alg}
\end{algorithm}
A practical consideration arises when generating a large number of channels using the CG procedure. In finite-precision computations, the theoretical orthogonality of the residual vectors is gradually degraded as the number of iterations increases, which can adversely affect the independence of the resulting channels. To mitigate this issue, the orthogonality between the newly generated channel and the previously computed channels is examined using the normalized inner product, and if a prescribed orthogonality criterion defined by a threshold $\tau_\text{orth}$ is not satisfied, a re-orthogonalization based on the modified Gram–Schmidt (MGS) process \cite{trefethen1997numerical} is employed to restore numerical orthogonality, as shown in Lines 6-10 in Algorithm \ref{alg}.

The mean-difference vector $\Delta \bar{\bar{\mathbf{g}}}$ is known for SKE tasks. For more complex signal detection tasks in which the signal is unknown, estimating $\Delta \bar{\bar{\mathbf{g}}}$ is needed. In such cases, 
prior information about object statistics, such as those corresponding to sparsity-promoting regularization and compressed sensing \cite{tibshirani1996regression, donoho2006compressed, sidky2008image, bora2017compressed}, may be useful.
To incorporate knowledge of background statistics,
the covariance matrix $\mathbf{K_g}$ can be estimated using covariance matrix decomposition (CMD) when a stochastic object model is available or has been learned.

\section{Numerical Studies}
\label{sect:num}
Numerical studies employing a stochastic object model were conducted to investigate the effectiveness of the proposed CG-based channels for binary detection tasks. A signal-known-exactly (SKE) and background-known-statistically (BKS) task was considered. 

\subsection{Objects and the imaging system}
A Type-I lumpy background model \cite{rolland1992effect} was employed to generate background images, in which the background objects were modeled as a superposition of ``lumps" randomly distributed over a $64\times 64$ (a.u.) field of view (FOV):
\begin{equation}
f_b(\mathbf{r}) = \sum_{n=1}^{N_b} l(\mathbf{r} - \mathbf{r}_n),
\end{equation}
where $N_b$ is the number of ``lumps", which is a Poisson random variable with mean $\overline{N}$, and $l(\mathbf{r} - \mathbf{r}_n)$ is the lump function centered at location $\mathbf{r}_n$. Here, $\mathbf{r} = (r_x, r_y)$ denotes a two-dimensional (2D) spatial coordinate. In this study, the $\overline{N} = 5$ and the lump function is Gaussian:
\begin{equation}
    l(\mathbf{r} - \mathbf{r}_n) = A \exp\left( -\frac{\|\mathbf{r} - \mathbf{r}_n\|^2}{2s^2} \right),
\end{equation}
where $A=1.2$, $s=7.8$, and $\mathbf{r}_n$ is sampled from a uniform distribution over the FOV of $64\times 64$.

To produce a more realistic signal with irregular shape, 
the signal was modeled as a mixture of Gaussian functions.
Specifically, the signal was simulated as a superposition of three Gaussian functions with different amplitudes, spatial locations, and widths:
\begin{equation}
s(\mathbf{r}) = \sum_{k=1}^{3} a_k \exp\!\left(
-\frac{(r_x - \mu_{k,x})^2}{2\sigma_{k,x}^2}
-\frac{(r_y - \mu_{k,y})^2}{2\sigma_{k,y}^2}
\right),
\end{equation}
where $\{a_k\}_{k=1}^3 = \{0.4, 0.28,0.32\}$, $\{\mu_{k,x}\}_{k=1}^3 = \{32, 37, 30\}$, $\{\mu_{k,y}\}_{k=1}^3 = \{32, 31, 36\}$, $\{\sigma_{k,x}\}_{k=1}^3 = \{3, 1.5, 1\}$, and $\{\sigma_{k,y}\}_{k=1}^3 = \{2, 1, 1.5\}$.

\begin{figure}[H]
    \centering
    \includegraphics[width=0.95\textwidth]{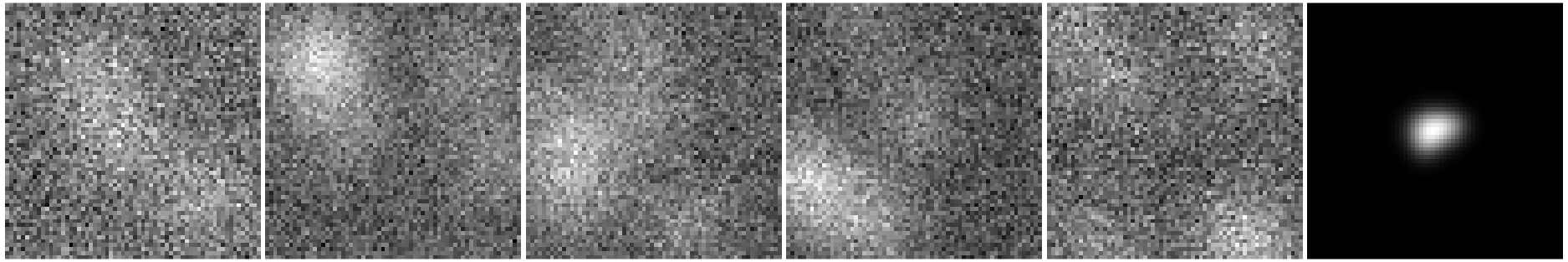}
    \vspace{0.2em}
    \caption{From left to right: five signal-present images and the signal image.}
    \label{fig:example}
\end{figure}

The imaging operator was modeled as a continuous-to-discrete (C-D) mapping with a Gaussian sensitivity function representing an idealized parallel-hole collimator system:
\begin{equation}
    h_m(\mathbf{r}) = \frac{h}{2\pi w^2} \exp\left( -\frac{(\mathbf{r} - \mathbf{r}_m)^T (\mathbf{r} - \mathbf{r}_m)}{2w^2} \right),
\end{equation}
where $h_m(\mathbf{r})$ describes the sensitivity of the $m$-th measurement to the object at location $\mathbf{r}$. Here, the width $w=2.5$ and the height $h=36$.
Because the background, signal, and sensitivity function were all modeled using Gaussian functions, the measured image could be computed analytically. \cite{zhou2019approximating}.
The measured images were simulated on a $64\times 64$ pixel grid. Examples of signal-present images and the signal image are shown in Fig. \ref{fig:example}.

\subsection{Implementation details}
The CG channels were generated using training datasets of varying sizes ranging from 400 to 20,000 images. Each dataset contained equal numbers of signal-absent and signal-present measured images and was employed to estimate the covariance matrix $\mathbf{K_g}$. The known signal image $\mathbf{s}$ served as the ground-truth mean-difference vector in the CG-based channel generation process.
The CG-CMD channels were generated in a manner similar to the CG channels, with the only difference being the estimation of the covariance matrix $\mathbf{K_g}$. Specifically, $\mathbf{K_g}$ was estimated through covariance matrix decomposition (CMD) using training datasets of background images together with the known noise covariance matrix, as described in Eq. (\ref{eq:K}). Different numbers of background images were considered for estimating $\mathbf{K_g}$ in the generation of the CG-CMD channels. 
For comparative evaluation, PLS channels were generated according to the procedure described by Witten~\emph{et al.} \cite{witten2010partial}. 

CHOs and CIOs were computed to perform the signal detection task. Specifically, the CHOs were computed by estimating the covariance matrix of the channelized data using a separate dataset containing 5,000 signal-absent images and 5,000 signal-present images. The HO estimated using CMD with 50,000 background images served as a reference for the CHOs. 
The CIOs and the corresponding IO reference were implemented using MCMC techniques \cite{kupinski2003ideal, park2007channelized}, with 200,000 steps and a burn-in period of 1,000.
Observer performance was evaluated using a test dataset comprising 500 signal-absent and 500 signal-present images.

The AUC value was employed to quantify the signal detection performance. The error bars associated with AUC values were estimated using bootstrapping with 1,000 bootstrap realizations. 
All model observer implementations and data analyses were performed in MATLAB R2025a on an Apple M4 Max chip.

\section{Results}
\label{sect:result}
\subsection{Channels and CHO templates}
\begin{figure}[H]
    \centering
    \begin{subfigure}{0.45\textwidth}
        \centering
        \includegraphics[width=\linewidth]{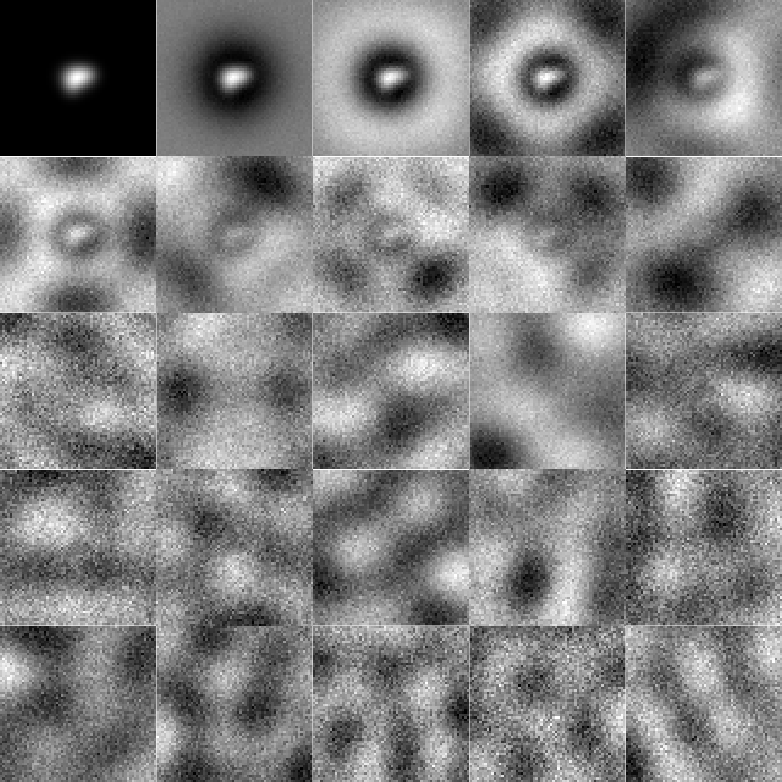}
        \caption{First 25 CG channels}
        \label{fig:sub1}
    \end{subfigure}
    \hfill
    \begin{subfigure}{0.45\textwidth}
        \centering
        \includegraphics[width=\linewidth]{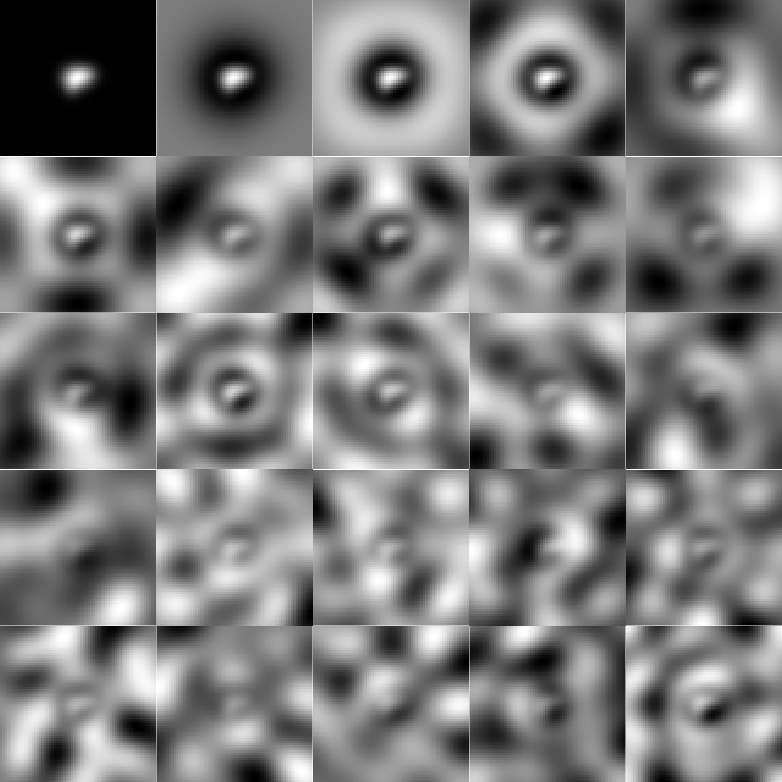}
        \caption{First 25 CG-CMD channels}
        \label{fig:sub2}
    \end{subfigure}
    \vspace{0.2em}
    \caption{CG and CG-CMD channels generated using 5000 images.}
    \label{fig:25channels}
\end{figure}

The first 25 CG and CG-CMD channels generated using 5,000 images are shown in Fig. \ref{fig:25channels} (a) and (b), respectively.
Both CG and CG-CMD produced channels that represented the signal appearance and the underlying statistical structure of the background, with CG-CMD produced cleaner channels due to CMD-based covariance matrix estimation using noise-free background images.

\begin{figure}[H]
    \centering
    \includegraphics[width=\textwidth]{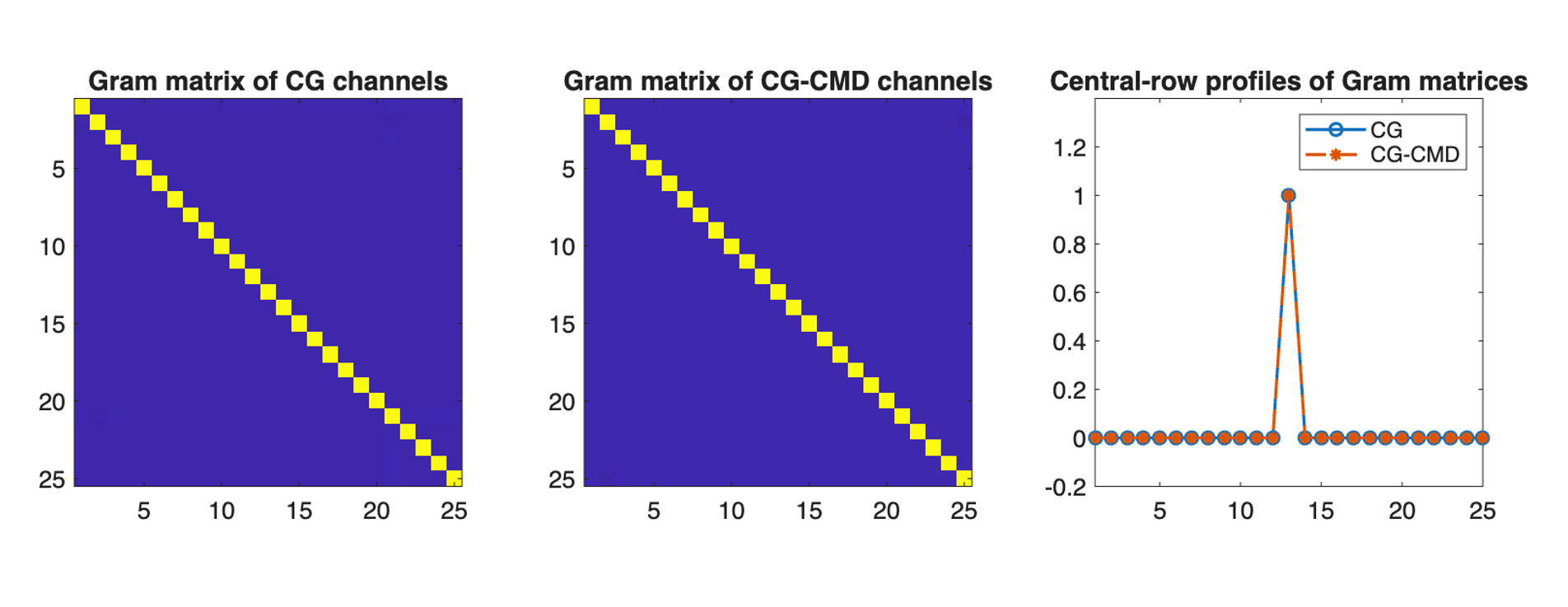}
    \caption{Gram matrices corresponding to the CG and CG-CMD channels.}
    \label{fig:gram}
\end{figure}

The Gram matrices corresponding to the 25 CG channels and 25 CG-CMD channels were evaluated to validate channel orthonormality. Each Gram matrix contained the pairwise inner products between channels, with the identity matrix corresponding to perfectly orthonormal channels. The resulting Gram matrices are shown in Fig. \ref{fig:gram}. As expected, both matrices were nearly identical to the identity matrix, indicating that the generated channels were approximately orthonormal.

\begin{figure}[H]
    \centering
    \includegraphics[width=\textwidth]{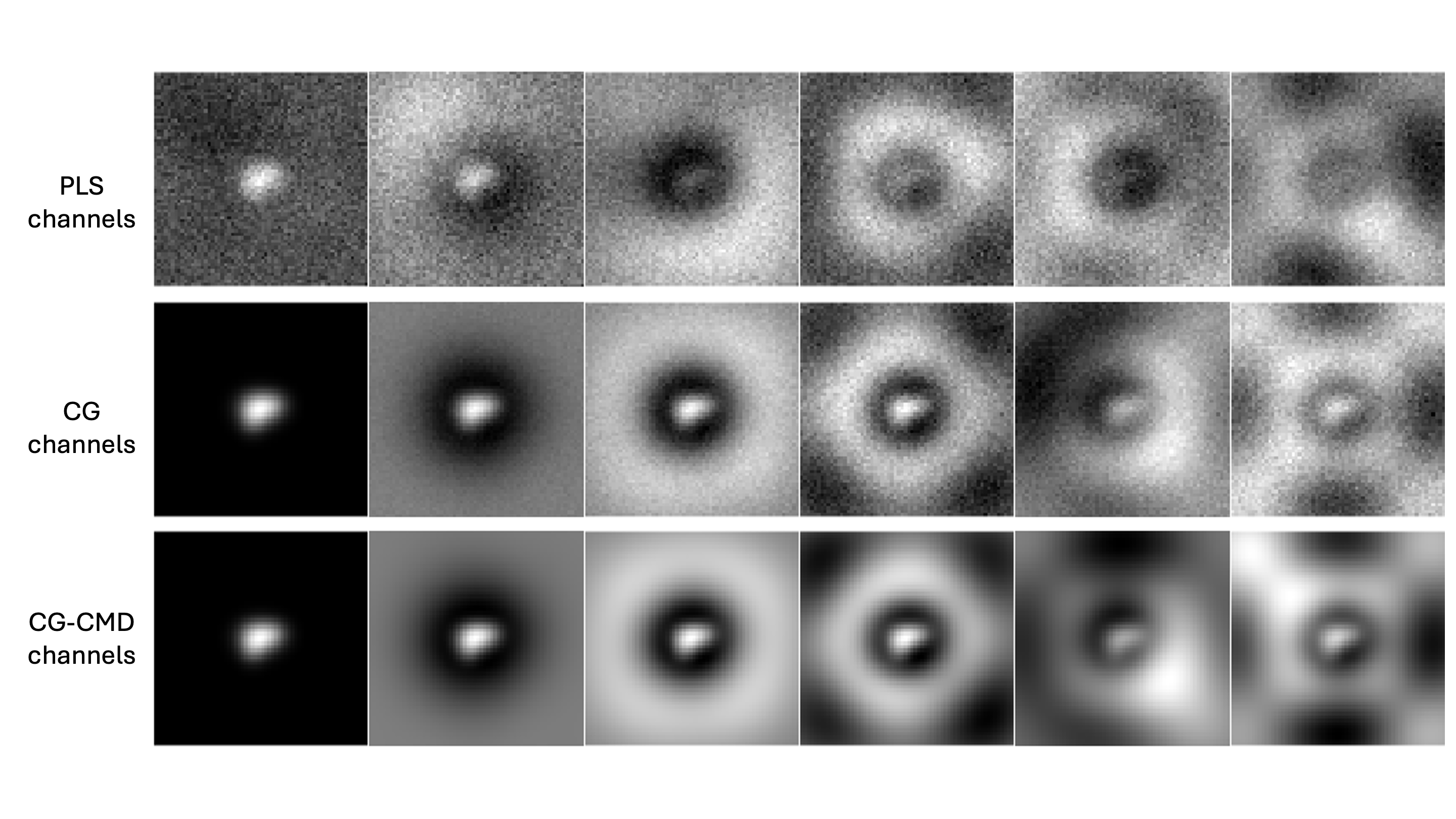}
    \caption{The first six PLS, CG, and CG-CMD channels.}
    \label{fig:6channels}
\end{figure}

The first six CG and CG-CMD channels are further compared with the PLS channels in Fig. \ref{fig:6channels}. 
Compared with the CG and CG-CMD channels, the PLS channels appeared substantially noisier.

\begin{figure}[H]
    \centering
    \includegraphics[width=\textwidth]{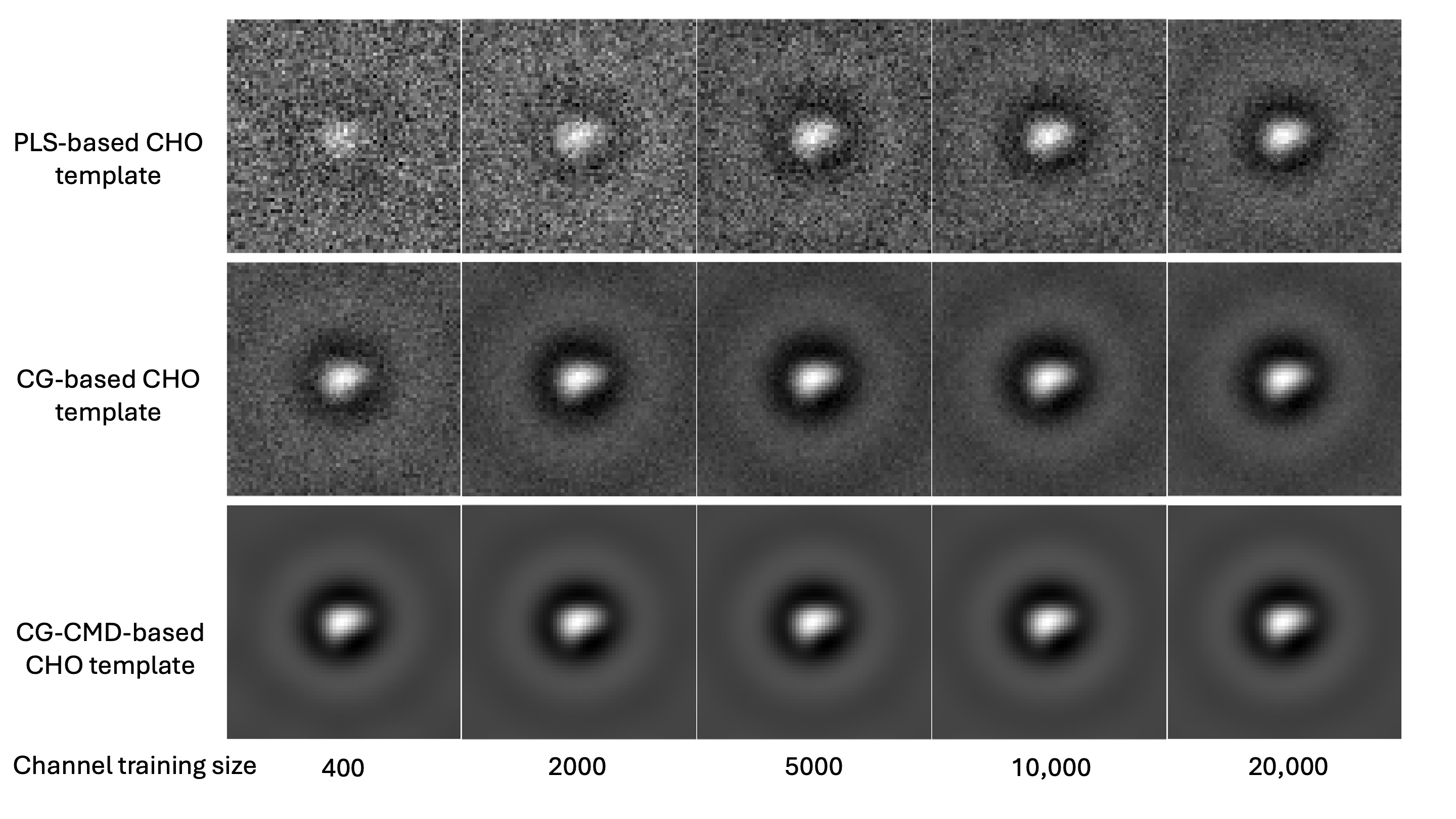}
    \caption{CHO templates computed from different channels with varying numbers of training images.}
    \label{fig:CHO_W}
\end{figure}

The CHO templates generated using different types of channels trained on varying numbers of images are shown in Fig. \ref{fig:CHO_W}. As expected, the templates computed from the PLS channels appeared the noisiest, whereas those generated from the CG channels had reduced noise, and those generated from the CG-CMD channels were the cleanest. The quality of the CHO template can significantly influence the signal detection performance, as demonstrated in the following subsection.

\subsection{Signal detection performance}
\begin{figure}[H]
    \centering
    \begin{subfigure}{0.495\textwidth}
        \centering
        \includegraphics[width=\linewidth]{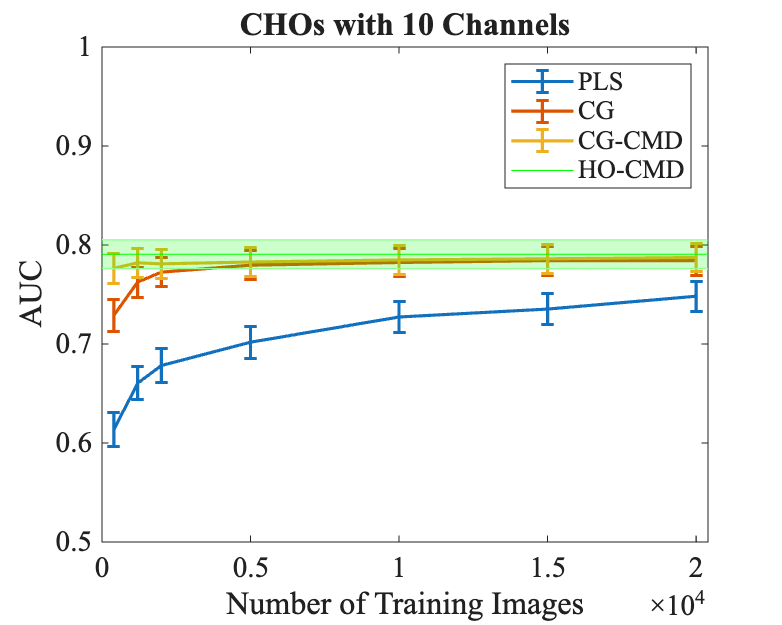}
        \caption{}
        \label{fig:sub1}
    \end{subfigure}
    \hfill
    \begin{subfigure}{0.495\textwidth}
        \centering
        \includegraphics[width=\linewidth]{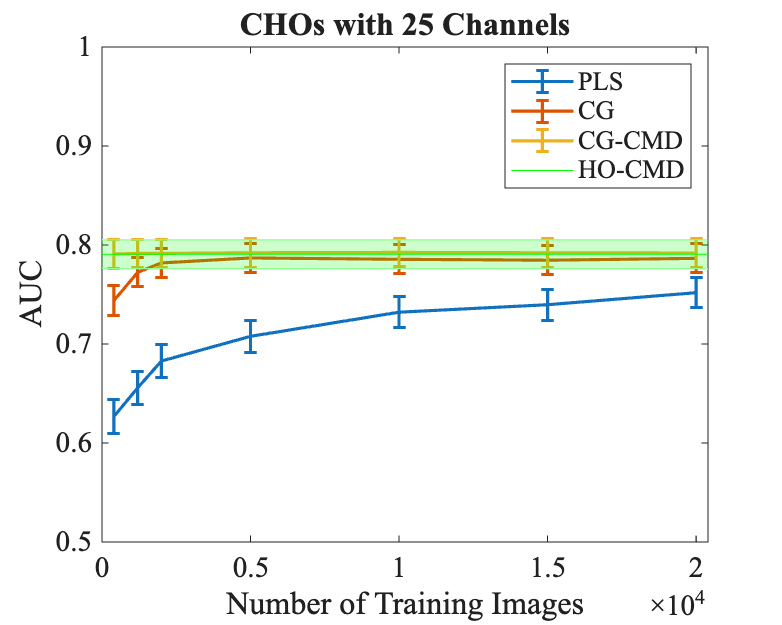}
        \caption{}
        \label{fig:sub2}
    \end{subfigure}
    \caption{AUC values of CHOs as functions of the number of training images for (a) 10 channels, and for (b) 25 channels.}
    \label{fig:AUC1}
\end{figure}

The AUC values corresponding to the CHOs with 10 and 25 channels, trained using different numbers of images ranging from 400 to 20,000, are shown in Fig. \ref{fig:AUC1}. 
The AUC values corresponding to the CG and CG-CMD channels were significantly higher than those obtained using the PLS channels. This is consistent with the observations in the preceding subsection, where the CG and CG-CMD channels were shown to be cleaner than the PLS channels.
When small training datasets were considered (e.g., 400 images), the AUC corresponding to the CHO with CG-CMD channels substantially exceeded that with the CG channels. Moreover, as shown in Fig. \ref{fig:AUC1} (b), the CHO employing 25 CG-CMD channels trained only using 400 images achieved detection performance that was nearly identical to that of the HO. These results demonstrate that the CG-CMD channels, which incorporate statistical properties of the background and noise, required only a small number of training images to construct efficient channels for the considered task.

\begin{figure}[H]
    \centering
    \begin{subfigure}{0.495\textwidth}
        \centering
        \includegraphics[width=\linewidth]{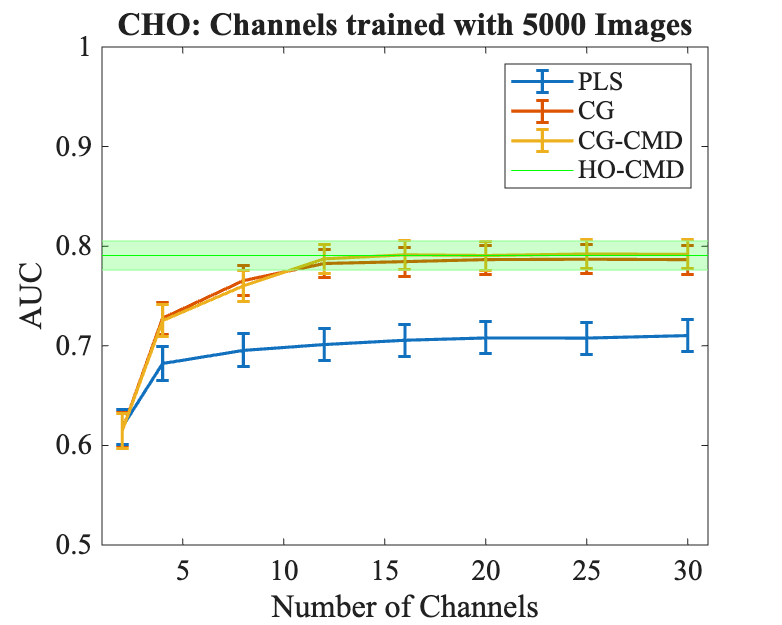}
        \caption{}
        \label{fig:sub1}
    \end{subfigure}
    \hfill
    \begin{subfigure}{0.495\textwidth}
        \centering
        \includegraphics[width=\linewidth]{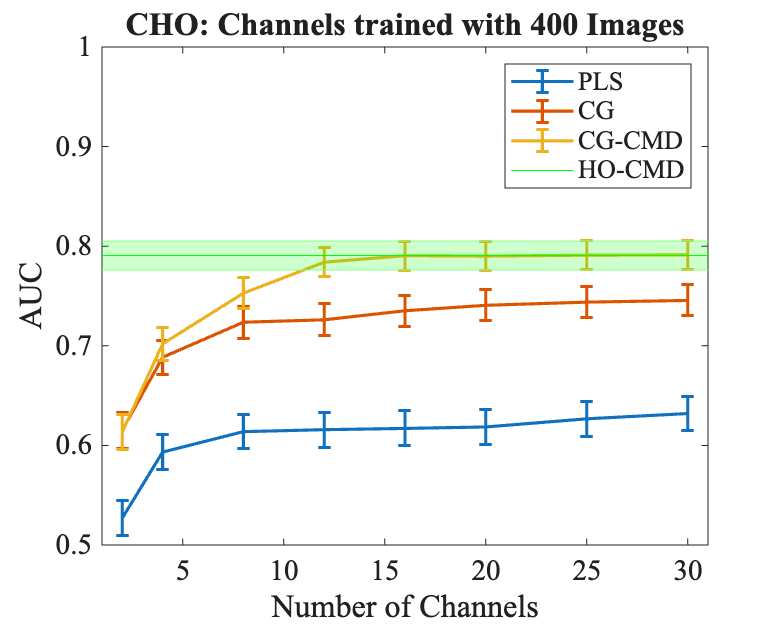}
        \caption{}
        \label{fig:sub2}
    \end{subfigure}
    \caption{AUC values of CHOs as functions of the number of channels for (a) 5,000 training images, and for (b) 400 training images.}
    \label{fig:AUC2}
\end{figure}

To further evaluate the signal detection performance corresponding to the considered channels under limited training data conditions,
CHOs employing different numbers of channels trained on datasets containing 5,000 and 400 images were computed.
The AUC values corresponding to the CHOs that employed varying numbers of channels trained on 5,000 images and 400 images are shown in Fig. \ref{fig:AUC2}. With 5,000 training images, CHOs employing CG channels and CG-CMD channels achieved very similar performance, and both substantially outperformed the CHO with PLS channels. When the number of training images was further reduced to 400, the differences in detection performance became more substantial. Neither the CG-based CHO nor the PLS-based CHO achieved the HO performance under such limited data condition, whereas the CG-CMD-based CHO still closely approximated the HO performance when 12 or more channels were employed.

\begin{figure}[H]
    \centering
    \begin{subfigure}{0.495\textwidth}
        \centering
        \includegraphics[width=\linewidth]{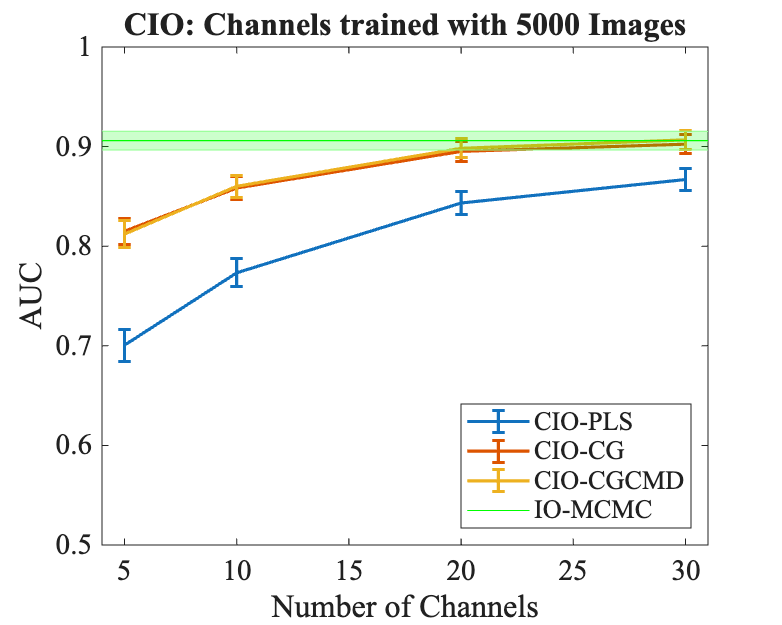}
        \caption{}
        \label{fig:sub1}
    \end{subfigure}
    \hfill
    \begin{subfigure}{0.495\textwidth}
        \centering
        \includegraphics[width=\linewidth]{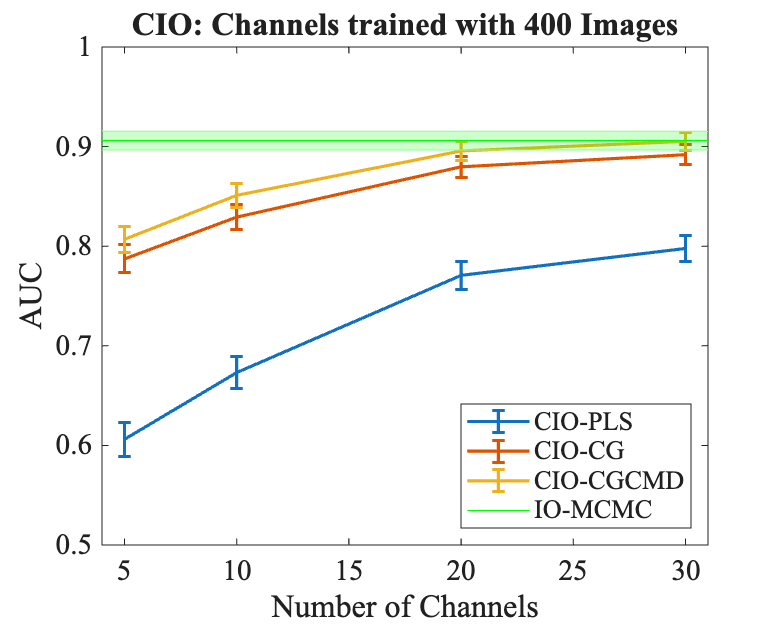}
        \caption{}
        \label{fig:sub2}
    \end{subfigure}
    \caption{AUC values corresponding to CIOs as functions of the number of channels for (a) 5,000 training images, and for (b) 400 training images.}
    \label{fig:CIO}
\end{figure}

The channelized Ideal Observer (CIO) was implemented using an MCMC algorithm \cite{park2007channelized}. The AUC values of the CIOs as functions of the number of channels for two training data sizes are shown in Fig.~\ref{fig:CIO}. The CIOs employing the CG and CG-CMD channels substantially outperformed those employing the PLS channels for all considered channel numbers under both training dataset conditions. With 5,000 training images, the CIOs with CG and CG-CMD channels performed nearly identically and both approximated the IO performance when 20 or more channels were used, whereas the CIOs employing PLS channels showed significant discrepancies relative to the IO performance. When only 400 training images were employed, the CIOs employing the CG-CMD channels outperformed both the CG-based and the PLS-based CIOs, and still closely approximated the IO performance when 30 channels were employed.

\section{Discussion and Conclusion}
\label{sect:conclusion}
In this work, we investigated the design of efficient channels based on the conjugate gradient (CG) method \cite{shewchuk1994introduction} for approximating ideal observers. 
The results demonstrated that the CG-based channels, including those constructed using covariance matrix decomposition (CMD), provided efficient low-dimensional image representations and produced superior signal detection performance compared with partial least squares (PLS) channels \cite{witten2010partial} for both CHO and CIO implementations across a range of training conditions.

Similar to the PLS-based channelization, the implementation of CG-based channels is not restricted to specific signals, backgrounds, or imaging operators. Moreover, CG-based channels possess an important advantage over PLS channels in that they can readily incorporate prior information about objects being imaged. The numerical studies demonstrated that incorporation of such object prior can significantly improve channel quality and signal detection performance, particularly under limited training data conditions. Such prior information may be available a priori (e.g., signals in signal-known-exactly tasks), estimated from data (e.g., covariance matrices estimated from synthesized background images via CMD), or designed to reflect specific modeling assumptions (e.g., sparsity in objects). An important direction for future work is to investigate how such prior information can be effectively utilized for more complex tasks and realistic imaging conditions. For example, when the signal is unknown, it may be estimated using methods that employ appropriately designed regularization.

Another important direction for future work is the integration of the proposed channelization framework with modern machine learning and deep learning methods \cite{kupinski2002ideal, zhou2019approximating, zhou2020approximating, li2021hybrid}. The proposed CG-based channels provide a principled approach for dimensionality reduction while preserving task-relevant information, and may therefore serve as an effective preprocessing step for downstream learning algorithms. Such low-dimensional representations may enable more efficient training and improved generalization, particularly under conditions where only limited training samples are available.

\subsection*{Disclosures}
The authors declare that there are no financial interests, commercial affiliations, or other potential conflicts of interest that could have influenced the objectivity of this research or the writing of this paper.

\subsection* {Code, Data, and Materials Availability} 
The data and code are available from the authors upon reasonable request. The numerical studies can be reproduced following the procedures described in Sec. \ref{sect:method} and the implementation details provided in Sec. \ref{sect:num}.

\subsection* {Acknowledgments}
The author gratefully acknowledges startup funding support provided by the Wyant College of Optical Sciences and the Department of Radiology \& Imaging Sciences at the University of Arizona. This work was inspired in part by the foundational contributions of Harrison H. Barrett to image science and the assessment of image quality.


\bibliography{report}   
\bibliographystyle{spiejour}   


\vspace{2ex}\noindent\textbf{Weimin Zhou} is an assistant professor at the University of Arizona, with dual appointments in the Wyant College of Optical Sciences and the Department of Radiology \& Imaging Sciences, and additional appointments in the Department of Biomedical Engineering and the Program in Applied Mathematics.
He is a recipient of the SPIE Community Champion Award and the SPIE Medical Imaging Cum Laude Award. His research lies at the intersection of image science and machine learning, with a focus on developing computational models and AI methods for assessing and improving image quality.


\end{spacing}
\end{document}